\documentclass{iaus}
\usepackage{graphicx}

\title[Binary evolution and evolutionary population synthesis] %% give here short title %%
{Evolution of binary stars and its implications for evolutionary population
synthesis}

\author[Han et al.]   %% give here short author list %%
{Z. Han$^1$, X. Chen$^1$, F. Zhang$^1$, 
Ph. Podsiadlowski$^2$}

\affiliation{$^1$National Astronomical Observatories / 
Yunnan Observatory, the Chinese Academy of Sciences, Kunming, 650011, China 
(email: zhanwenhan@ynao.ac.cn)
\\[\affilskip]
$^2$University of Oxford, Department of Astrophysics, Keble Road, 
Oxford, OX1 3RH, UK}

\pubyear{2009}
\volume{262}  %% insert here IAU Symposium No.
\pagerange{1--4}
\jname{Stellar Populations: Planning for the Next Decade}
\editors{G. Bruzual \& S. Charlot, eds.}
\begin{document}

\maketitle

\begin{abstract}
Most stars are members of binaries, and the evolution of a star 
in a close binary system differs from that of an ioslated star 
due to the proximity of its companion star. The components in 
a binary system interact in many ways and binary evolution 
leads to the formation of many peculiar stars, including blue 
stragglers and hot subdwarfs. We will discuss binary evolution 
and the formation of blue stragglers and hot subdwarfs,
and show that those hot objects are important in the study of 
evolutionary population synthesis (EPS), and conclude that 
binary interactions should be included in the study of
EPS. Indeed, binary interactions make a stellar population 
younger (hotter), and the far-ultraviolet (UV) excess in 
elliptical galaxies is shown to be most likely 
resulted from binary interactions. This has major 
implications for understanding the evolution of 
the far-UV excess and elliptical galaxies in general. 
In particular, it implies that the far-UV excess is 
not a sign of age, as had been postulated prviously and predicts
that it should not be strongly dependent on the 
metallicity of the population, but
exists universally from dwarf ellipticals to giant ellipticals.

\end{abstract}

\firstsection % if your document starts with a section,
\section{Introduction}

Evolutionary population synthesis (EPS) has experienced a 
rapid progress since the early 90's and provides the most robust
approach in studying stellar populations of galaxies. In most
of the current EPS models, binary evolution has been ignored. 
However, most stars are members of binaries, and binary evolution 
leads to the formation of many peculiar objects, such as blue stagglers and 
hot subdwarfs, which are hot, long-lived and still very luminous.
Those objects contribute very much to the spectral energy distribution
(SED) at short wavelength for an old stellar population, and make
the population look hotter and younger. Such a ``cosmetic'' effect 
has been successfully included by EPS models of 
\cite[Zhang et al.\ (2004)]{zha04}, 
\cite[Han, Podsiadlowski \& Lynas-Gray (2007)]{han07},
\cite[Chen \& Han (2009)]{che09}.
Some puzzles in EPS
have been solved with the inclusiong of binaries and binary
evoltuion is an important ingredient in EPS and also a subject of this
symposium.

\section{Binary evolution}
Most stars are in binaries and binaries evolve differently from 
single stars. Binary evolution is much more complicated and produces
many interesting objects, such as cataclysmic variables, type Ia supernovae,
X-ray binaries, symbiotic stars, blue stragglers, hot subdwarfs, etc. 
Binary evolution leads to rejuvenation of stars and
the formation of hotter objects in an otherwise old cold population.

There are numerous evolutionary channels for a binary system, depending
on its initial conditions. However, the main evolutionary processes
can be described as below. A binary system (of low/intermediate mass) 
has two components: the primary (the initially more massive
one) and the secondary. As the binary evolves, the primary expands and 
may fill its Roche lobe, and Roche lobe overflow (RLOF) begins and
mass is transferred from primary to secondary.
If RLOF is dynamically stable, the mass transfer removes the primary's
envelope and its hot core is exposed, and in some cases the core can
be ignited (i.e. long-lived). The secondary grows in mass via accretion and 
this rejuvanates the secondary. Stable RLOF leads to the formation of
a wide binary system.
If the mass transfer is dynamically unstable, the mass transfer leads to
the formation of a common envelope (CE).
The CE engulfs the core of the primary and the secondary, and
does not co-rotate with the embedded binary. The friction between the CE
and the embedded binary makes the orbit decay, and a large amount of
orbital energy released is deposited into the CE. If the CE can be ejected,
a close binary forms with the primary's core being exposed,
otherwise the CE evolution leads to the formation of
a fast rotating merger, which is a more massive and hotter star.

As shown above, a binary may produce hotter objects in the following three 
ways. 1) The envelope of a component is removed via stabel RLOF or CE ejection
and leaves a naked (sometime burning) hot core. 2) A star grows in mass
via accretion due to stable RLOF and gets rejuvenated. 3) Coalescence of a
binary system produces a more massive and hotter star.

\section{Blue stragglers and the contribution to SED}
Blue straggers (BSs) are stars that have remained on the main 
sequence for a time exceeding that expected, for their masses, 
from standard stellar evolution theory. 
BSs are blue and luminous, and may contribute significant excess spectral 
energy in the blue and ultraviolet of SED of a stellar population.
Those objects are relevant to primordial binaries. Binary
coalescence from a contact binary is a major channel
for single BSs and it is accepted that the contact binaries
are mainly from case A mass transfer, which is defined as the primary 
being on the main sequence at the onset of mass transfer.
Another channel to produce BSs from primordial binaries
is mass transfer, where the secondary accretes some material
and grows in mass. The primary can be on the the main
sequence (case A) at the onset of mass transfer, or after
the main sequence but before central He burning (case B),
or during or after central He burning (case C).

\begin{figure}
\centerline{
\includegraphics[height=10.0cm,angle=270]{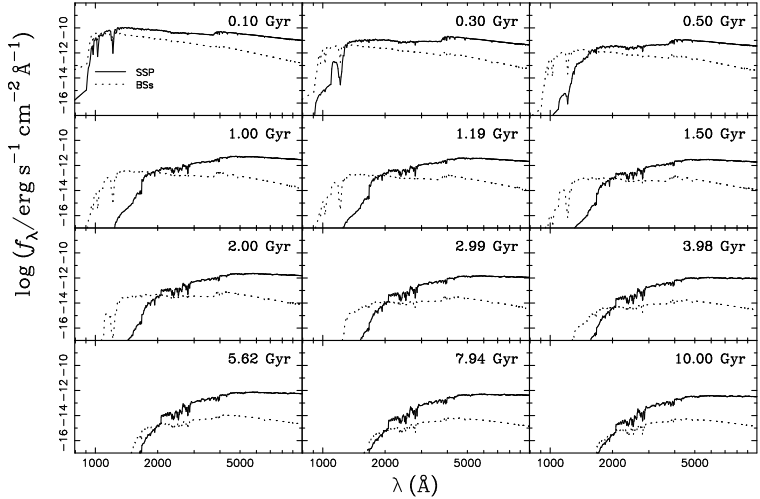}
}
\caption{
The evolution of SED for a SSP without binary interactions considered
(solid line) and the contribution of BSs (dotted line)
}
\label{bs}
\end{figure}

Fig.~\ref{bs} shows an integrated rest-frame intrinsic SEDs at different
ages for a simple stellar population (including binaries) 
with a mass of $10^{11}M_\odot$ at a
distance of 10Mpc. The solid lines are for the
results without binary
interactions considered, and the dotted ones are the contribution 
of BSs from primordial binary evolution, 
i.e mass transfer and coalescence of two components 
(\cite[Chen \& Han 2008a, 2008b, 2009]{che08a,che08b,che09}).

\section{Hot subdwarfs and the UV-upturn phenomenon of elliptical galaxies}

Hot subdwarfs 
(including subdwarf B stars, subdwarf O stars and subdwarf OB stars), 
known as extreme horizontal branch (EHB) stars in globular 
clusters, are generally considered
to be helium-core-burning stars with extremely thin hydrogen envelopes
($<0.02M_\odot$).
The leading theory for their formation is the so called
binary model proposed by 
\cite[Han et al.\ (2002, 2003, 2008)]{han02,han03,han08}, which 
successfully explained the main observational 
characteristics of field hot subdwarfs and EHB
stars in globular clusters.
It is now well established that the vast
majority of (and quite possibly all) Galactic hot
subdwarfs and EHB stars in globular clusters
are the results of binary interactions, where a star loses
all of its envelope near the tip of the red-giant branch by mass
transfer to a companion star or ejection in a common-envelope phase,
or where two helium white dwarfs merge with a combined mass
larger than $\sim 0.35\,M_{\odot}$ (see 
\cite[Han et al. 2002, 2003, 2008]{han02,han03,han08} 
for references and details). In all of these cases, the remnant star 
{\it ignites helium} and becomes a hot subdwarf. 
{\it The key feature of these binary
channels is that they provide the missing physical mechanism for
ejecting the envelope and producing a hot subdwarf.}
As we will show below, the binary hot subdwarf model provides a natural
explanation for the UV-upurn phenomenon of elliptical galaxies.

The UV-upturn phenomenon is a long-standing problem in the study 
of elliptical galaxies, where there exists a
far-ultraviolet (UV) excess in their spectra. 
It is now clear that UV-upturn is caused by hot subdwarfs.
Two scenarios, referred to as the high- and the low-metallicity scenario, 
had been advanced (\cite{yi97}; \cite{lee94}) before the binary model
was invented. In those scenarios,
a star loses all of its envelope via stellar wind near the tip of the red
giant branch and the naked core gets ignited and becomes a hot subdwarf.
These scenarios are {\it ad hoc} (see 
\cite[Han, Podsiadlowski \& Lynas-Gray 2007]{han07} for details)
and require a large age for the hot subdwarfs and therefore predict
that the UV excess declines rapidly with redshift. This is not
consistent with recent observations, e.g. with the Hubble Space
Telescope (HST) (\cite[Brown et al. 2003]{bro03}). 

\begin{figure}
\centerline{
\includegraphics[height=9.5cm,angle=270]{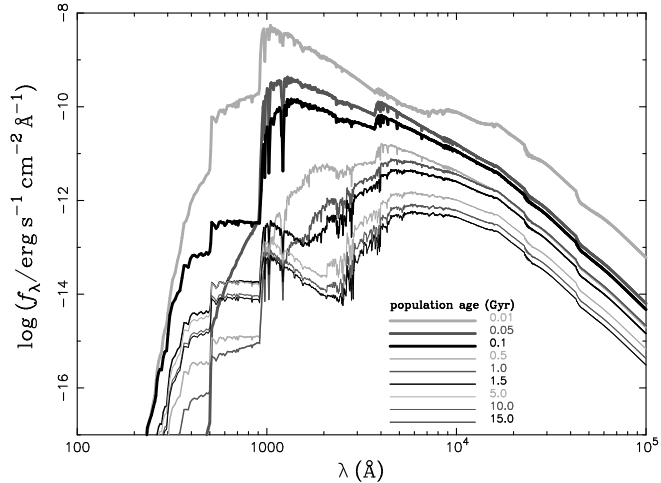}
}
\caption{
The evolution of the rest-frame intrinsic SED for a simulated galaxy
in which all stars formed at the same time, i.e. a simple stellar population
(SSP).  The stellar population (including
binaries) has a mass of $10^{11}M_\odot$ and the galaxy is assumed to
be at a distance of 10\,Mpc.  
}
\label{sed}
\end{figure}

The binary model naturally expained the formation of hot subdwarfs, and
it would be {\it a priori} to apply the binary model of hot subdwarfs
to the UV-upturn problem 
(\cite[Han, Podsiadlowski \& Lynas-Gray 2007]{han07}).
Figure~\ref{sed} shows our simulated evolution of the SED 
of a galaxy in which all the stars formed at the
same time.  At early times, the far-UV flux is entirely
caused by the contribution from young stars. Hot subdwarfs from the
various binary evolution channels become important after about
1.1\,Gyr, and soon start to dominate completely. After a
few Gyr the far-UV SED no longer changes appreciably relative to the
visual flux. One immediate implication of this is that the model
predicts that the magnitude of the UV excess $(1550-V)$, defined as
the relative ratio of the flux in the $V$ band to the far-UV flux, 
should not evolve significantly with look-back time or
redshift. Indeed, this is exactly what seems to have been found in
recent observations  
(\cite[Brown et al. 2003, Atlee, Assef \& Kochanek 2009]{bro03,atl09}). 

Despite its simplicity, our model can successfully reproduce most of
the properties of elliptical galaxies with a UV excess: e.g., the
range of observed UV excesses, both in $(1550-V)$ and $(2000-V)$, 
and their evolution with redshift.  The model shows
that the UV excess is not a
good age indicator, as has been argued previously,
and that all old galaxies should show a UV
excess at some level. Moreover, we expect that the model is not very
sensitive to the metallicity of the population since metallicity does
not play a significant role in the envelope ejection process.

Using the binary UV-upturn model, 
\cite[Lisker \& Han (2008)]{lis08} studied
the GALEX far- near-UV colours of Virgo Cluster
early-type galaxies, and found that, 
consistent with galaxy formation theories,
the opposite behavior in the colours of dwarf ellipticals and giant 
ellipticals naturally occurs without the requirement of a dichotomy 
between the stellar populaiton properties of dwarfs and giants, in contrast 
to previous conclusions.

%\section{Conclusions}
%Binary evolution plays a crucial role in SED at short wavelength
%for an old, passively evovling stellar population.
%
%\section{Acknowledgements}
This work is supported by the Natural Science Foundation of China under
grant nos. 10821061, 10973036, 10773026 and 2007CB815406.

\end{document}